\documentstyle[aps,pra,twocolumn,floats,epsfig]{revtex}

\begin{document}

\renewcommand{\floatpagefraction}{0.8}
\setlength{\textfloatsep}{2pt plus 1pt minus 1pt} 

\draft

\def\overlay#1#2{\setbox0=\hbox{#1}\setbox1=\hbox to \wd0{\hss #2\hss}#1%
\hskip
-2\wd0\copy1}
\twocolumn[
\hsize\textwidth\columnwidth\hsize\csname@twocolumnfalse\endcsname

\title{High-order nonlinearities in the motion of a trapped atom}

\author{S. Wallentowitz and W. Vogel}

\address{Arbeitsgruppe Quantenoptik, Fachbereich Physik, Universit\"at
  Rostock, \\ Universit\"atsplatz 3, D--18051 Rostock, Germany}

\author{P.L. Knight}

\address{Blackett Laboratory, Imperial College, London SW7 2BZ, United
  Kingdom}

\date{June 10, 1998}

\maketitle

\begin{abstract}
  We study the counterpart to the multi-photon down conversion in the
  quantised motion of a trapped atom. The Lamb--Dicke approximation
  leads to a divergence of the mean motional excitation in a finite
  interaction time for $k$-quantum down conversions with $k\!\geq\!3$,
  analogous to the situation in the parametric approximation of
  nonlinear optics.  We show that, in contrast to the Lamb--Dicke
  approximation, the correct treatment of the overlap of the atomic
  center-of-mass wave function and the driving laser waves leads to a
  proper dynamics without any divergence problem.  That is, the wavy
  nature of both matter and light is an important physical property
  which cannot be neglected for describing the motional dynamics of a
  trapped atom, even for small Lamb--Dicke parameters.
\end{abstract}

\pacs{PACS numbers: 42.50.Vk, 32.80.Lg, 42.65.-k, 03.65.-w} \vskip2pc]

\narrowtext

\section{Introduction}
When the susceptibility of a medium interacting with an
electromagnetic field of optical frequency depends strongly on the
electric-field amplitude, one enters the domain of nonlinear optics.
Nonlinear couplings of electric fields of different frequencies
usually emerge from an expansion of the susceptibility in terms of the
electric-field amplitude. Prominent examples of such nonlinear
couplings are second-harmonic generation or two-photon down
conversion, which are due to a second-order susceptibility $\chi_2$.
Nonlinear crystals have been successfully used to produce squeezed
quantum states of light via a two-photon down conversion.  The
extension of two-photon down conversion to an arbitrary $k$-photon
process, where $k\!>\!2$, has also been studied.  While this might be
viewed as a natural generalisation of the second-harmonic generation
or the two-photon down conversion, it has been shown that there is a
subtle problem in the theoretical description of such processes.
Fisher, Nieto, and Sandberg~\cite{nieto} have argued that it is not
possible to define states by applying the unitary time-evolution
operator on the vacuum field state. This argument has been partially
removed by a consideration using Pad{\'e}
approximants~\cite{braunstein}.  Later on, however, it was shown by
Elyutin and Klyshko~\cite{elyutin} and Hillery~\cite{hillery} that for
$k\!=\!3$ and $k\!=\!4$, respectively, a divergence occurs in the mean
photon number for finite interaction times. This divergence property
may be interpreted as an unphysical artefact coming from the improper
treatment of the $k$-photon process.  In fact, it has been shown that
the usual parametric approximation is incorrect in that it neglects
the energy transfer and entanglement, between the pump and signal mode
of the electromagnetic field which emerges when the pump mode is
quantised~\cite{buzek-drobny,drobny,drobny-buzek,banaszek-knight}.  We
note that the possibility of observing $k\!=\!3$ nonlinear quantum
optical conversion processes in a damped cavity has recently been
discussed~\cite{schiller}.

While these phenomena are well known and elaborated in the context of
nonlinear optics, recent advances in
laser-cooling~\cite{laser-cooling1,laser-cooling2}, state
preparation~\cite{wineland-number-squeeze,wineland-cat}, and
detection~\cite{leibfried} of the motional quantum state of single
trapped ions, a new type of realisation of such nonlinear mode
couplings became possible.  Here the modes are represented by the 3D
harmonic center-of-mass oscillations of a single ion in the trap.  The
nonlinear mode coupling may be realized by appropriate laser
irradiation which induces vibrational Raman
transitions~\cite{counterpart,agarwal,steinbach,drobny-fake,squeezing}.
This opens possibilities to study such mode couplings with an almost
perfect system (i.e. the motion of the trapped ion) where the damping
of the motion is negligibly small apart from a small heating rate due
to technical imperfections~\cite{wineland-review}. In the Lamb--Dicke
regime, where the atomic center-of-mass position is well localised
with respect to the wavelengths of the applied laser fields, mode
couplings result which are analogous to the optical mode couplings in
the parametric approximation. That is, a treatment of the dynamics
based on the Lamb--Dicke approximation would reveal an unphysical
divergence of the mean number of vibrational quanta for $k$-quantum
processes with $k\!\geq \! 3$.

Whereas for a trapped atom in the Lamb--Dicke regime one gets a close
connection to the parametrically approximated optical couplings, for a
trapped atom which is not well localised with respect to the laser
wavelengths, nonlinear modifications of the couplings occur which
arise from the overlap of the atomic center-of-mass wavefunction with
the laser waves, describing the momentum transfer onto the atomic
center-of-mass during laser-photon absorption and
emission~\cite{counterpart,squeezing}.  These recoil effects may
strongly influence the coupling strengths which depend on the number
of excited quanta in the vibrational modes.  They have been
predicted~\cite{njcm} and observed~\cite{wineland-number-squeeze} for
the first time in the context of a nonlinear Jaynes--Cummings model,
describing the dynamics of a laser-driven vibronic transition in the
resolved-sideband regime.

We will show in this paper that the nonlinear effects, caused by the
overlap of light and matter waves, will remove the unphysical
divergence problem which arises in the Lamb--Dicke approximation.  The
paper is structured as follows: In Sec.~II we introduce the effective
Hamiltonian for the motional dynamics of the trapped atom and we
briefly discuss the justification and the validity of the parametric
approximation for optical couplings and the Lamb--Dicke approximation
for the motional couplings. The laser-driven motional dynamics is then
considered in Sec.~III and the divergence problem in the Lamb--Dicke
approximation is studied in Sec.~IV. In Sec.~V the more realistic
treatment of an unspecified degree of localisation of the trapped atom
is shown to remove the divergence and some examples for the time
evolution are given. A summary and some conclusions are found in
Sec.~VI.

\section{Motional counterpart of multi-photon down conversion}
\begin{figure}
  \begin{center}
    \psfig{file=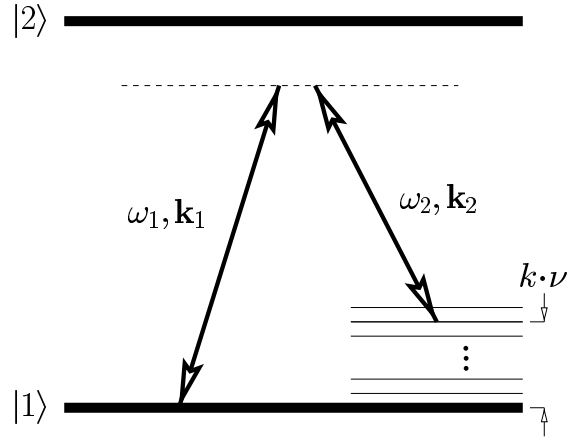,scale=0.6}
    \vspace*{2ex}
    \caption{$k$-quantum motional coupling by application
      of two off-resonant laser fields with laser difference-frequency
      $\omega_1 \!-\! \omega_2 \!=\! k \nu$, where $\nu$ is the
      frequency of the vibrational mode which is specified by the
      beam-directions, ${\bf k}_1 \!-\! {\bf k}_2$, of the beat node
      of the two lasers.}
    \label{fig:excitation-scheme}
  \end{center}
\end{figure}

For the $k$-quantum vibrational-mode coupling, we consider here a
two-photon vibrational Raman transition which has been experimentally
realised~\cite{wineland-number-squeeze,wineland-cat} and theoretically
studied in the context of nonlinear couplings of vibrational
modes~\cite{counterpart,agarwal,steinbach,drobny-fake}. By application
of two laser beams which are off-resonant with respect to a strong
electronic dipole transition and which are detuned relative to each
other by multiples of the vibrational frequencies in the trap,
vibrational Raman transitions can be driven which may be used to
realize a quantum mechanical counterpart of nonlinear optics, see
Fig.~\ref{fig:excitation-scheme}.  
\noindent
For an appropriate laser-beam propagation geometry which affects only
the dynamics in one vibrational mode of frequency $\nu$, in the
rotating-wave approximation the Hamiltonian describing the effect of
the Raman laser drive on the dynamics of the vibrational mode is given
by~\cite{counterpart}
\begin{equation}
  \label{eq:k-hamiltonian}
  \hat{H}_k = \hbar \kappa \, \hat{f}_k(\hat{a}^\dagger \hat{a}; \eta)
  \, (i\eta\hat{a})^k + {\rm h.c.}
\end{equation}
where $\hat{a}$ and $\hat{a}^\dagger$ are the annihilation and
creation operators of vibrational quanta, respectively. Here the laser
difference-frequency has been chosen to be $k \nu$, i.e.  $k$ times
the vibrational frequency of the mode with
$k\!\geq\!1$~\cite{k-remark}.  The effective two-photon coupling
strength is given by $\kappa$, and
\begin{displaymath}
  \eta \!=\! 2 \pi \, \frac{\sqrt{\langle 0 | {\Delta \hat{x}}^2 | 0
      \rangle}}{\lambda} ,
\end{displaymath}
is the so-called Lamb--Dicke parameter describing the localisation of
the spatial extension of the center-of-mass wavefunction in the ground
state relative to the wavelength $\lambda$ of the beat-node of the two
laser beams. The Hermitian operator functions
$\hat{f}_k(\hat{a}^\dagger \hat{a}; \eta)$ strongly depend on $\eta$
and are defined by the normally ordered expressions
\begin{eqnarray}
  \label{eq:k-function}
  \hat{f}_k(\hat{a}^\dagger \hat{a}; \eta) & = & e^{-\eta^2/2}
  \sum_{l=0}^\infty (-1)^l \frac{\eta^{2l}}{l! \, (l+k)!} \,
  \hat{a}^{\dagger l} \hat{a}^l \\ \nonumber & = & : \, \left( 2 \eta
    \sqrt{\hat{a}^\dagger \hat{a}} \right)^{-k} J_k\left( \eta
    \sqrt{\hat{a}^\dagger \hat{a}} \right) \, e^{-\eta^2/2} \, : .
\end{eqnarray}
These nonlinear operator functions correspond to overlap integrals of
the motional states $|n\rangle$, $|n\!+\!k\rangle$ and the beat-node
of the laser fields. They account for the recoil effects during the
process of absorption and emission of laser photons of the trapped
atom.  Since they depend only on the quantum number $\hat{n} \!=\!
\hat{a}^\dagger \hat{a}$, in the basis of its eigenstates,
$\hat{n}|n\rangle \!=\! n |n\rangle$ ($n\!=\! 0,1,2,\ldots$), these
operators are diagonal, with their diagonal elements $f_k(n;\eta)
\!=\!  \langle n | \hat{f}_k(\hat{a}^\dagger \hat{a}; \eta) | n
\rangle$ being given by
\begin{equation}
  \label{eq:k-func-fock}
  f_k(n;\eta) = \frac{n!}{(n+k)!} \, L_n^{(k)}(\eta^2) \,
  e^{-\eta^2/2} ,
\end{equation}
where $L_n^{(k)}(x)$ are the associated Laguerre polynomials.  For a
well-localised atom, that is, for very small Lamb--Dicke parameters
$\eta \!\ll\! 1$, or more precisely for a small spatial extension of
the atomic wavefunction $\eta\sqrt{n+1} \!\ll\! 1$, one reaches the
so-called Lamb--Dicke limit. Here usually the Lamb--Dicke
approximation is made, which takes into account only the lowest-order
terms in $\eta$. In our description of the $k$-quantum
coupling~(\ref{eq:k-hamiltonian}) the Lamb--Dicke approximation is
performed by replacing the operator-valued function
$\hat{f}_k(\hat{a}^\dagger \hat{a};\eta)$ by its limiting value for a
small Lamb--Dicke parameter,
\begin{equation}
  \label{eq:k-func-limes}
  \lim_{\eta\to 0} \hat{f}_k(\hat{n}; \eta) = \frac{1}{k!} .
\end{equation}
By replacing in the Hamiltonian~(\ref{eq:k-hamiltonian}) the operator
function $\hat{f}_k(\hat{n}; \eta)$ by the c-number given in
Eq.~(\ref{eq:k-func-limes}) one obtains the interaction Hamiltonian
$\hat{H}_k^{\rm (LD)}$ of the $k$-quantum process in the Lamb--Dicke
approximation, that is, in lowest order of the Lamb--Dicke parameter,
\begin{equation}
  \label{eq:k-hamilton-limes}
  \hat{H}_k^{\rm (LD)} = \hbar \kappa_k \, \hat{a}^k + \hbar
  \kappa_k^\ast \, \hat{a}^{\dagger k} ,
\end{equation}
with 
\begin{displaymath}
  \kappa_k = \kappa \, \frac{(i\eta)^k}{k!}
\end{displaymath}
being the $k$-quantum coupling strength in the Lamb--Dicke
approximation.

In the context of nonlinear optics the
Hamiltonian~(\ref{eq:k-hamilton-limes}) describes the $k$-photon down
conversion process where $\hat{a}$ is the signal mode and the pump
mode has been parametrically approximated by replacing its operators
by amplitudes whose values are included in the coupling strength
$\kappa_k$.  It is well known that for $k\!>\!2$ the parametric
approximation described by the Hamiltonian~(\ref{eq:k-hamilton-limes})
leads to a divergent behaviour of the mean quantum number in the
signal mode for {\em finite} interaction times, i.e.
\begin{equation}
  \label{eq:divergence}
  \lim_{t_2-t_1 \to \Delta t_\infty} \langle \hat{n}(t_2) \rangle -
  \langle \hat{n}(t_1) \rangle = \infty ,
\end{equation}
for a defined interaction time $0 \!<\! \Delta t_\infty \!<\! \infty$.
In nonlinear optics the parametric approximation of the pump mode
fails due to pump depletion and the entanglement of signal and pump
modes which is essential in this type of interaction, leading to a
energy conservation of the total number of photons in the pump {\em
  and} signal mode~\cite{nieto}. The parametric approximation
essentially neglects the energy transfer from the signal to the pump
mode, leading to an unbounded increase of the energy in the signal
mode. A quantum description of the pump mode is therefore required,
regardless of how strong the pump field actually is compared with the
signal mode~\cite{buzek-drobny,drobny,drobny-buzek,banaszek-knight}.

For the case of a single trapped atom, the coupling strength
$\kappa_k$ contains the classically approximated field amplitudes of
the two Raman lasers. While in nonlinear optics the parametric
approximation for the pump mode fails it is expected that for a
Raman-driven trapped atom it is the Lamb--Dicke approximation, rather
than the replacement of the laser-field operators by their classical
amplitudes, which leads to a divergent dynamics. Note that the
validity of the Lamb--Dicke approximation is, in principle, in
contradiction with a divergent motional excitation in the trap.
Eigenstates of the trap potential should only be populated for $\eta
\sqrt{n+1} \!\ll\!  1$, that is for higher excitations the
approximation is no longer valid, and a diverging mean excitation
violates this requirement.

Therefore in all cases, even for a trapped atom with small Lamb--Dicke
parameter ($\eta \!\ll\! 1$), we have to consider the full problem
including the nonlinear operator functions $\hat{f}_k(\hat{n}; \eta)$
in the Hamiltonian~(\ref{eq:k-hamiltonian}). As already noted, these
operator functions describe the effects of momentum transfer onto the
atomic center-of-mass motion during the laser-atom
interaction~\cite{counterpart,squeezing,njcm}. They are of particular
importance for higher vibrational excitations $\eta \sqrt{n+1} \!>\!
1$ and they are discarded in the Lamb--Dicke approximation.

\section{Equations of motion}
To study the time evolution of the mean quantum number $\langle
\hat{n}(t) \rangle$, we will start by deriving from the
Hamiltonian~(\ref{eq:k-hamiltonian}) the equations of motion for the
populations of the vibrational levels. The Schr\"odinger equation is
given by
\begin{equation}
  \label{eq:schroedinger}
  i\hbar \frac{\partial}{\partial t} \, |\psi(t)\rangle = \hat{H}_k \,
  |\psi(t)\rangle ,
\end{equation}
where $|\psi(t)\rangle$ is the state vector in the interaction
picture. Using as basis the number states $|n\rangle$, we get the
following equations of motion for the coefficients $\psi_n(t) \!=\!
\langle n | \psi(t) \rangle$
\begin{equation}
  \label{eq:diff-eq1}
  \frac{\partial\psi_n}{\partial t} = -i \left[ g_k(n;\eta) \,
    \psi_{n+k} + g_k^\ast(n\!-\!k;\eta) \, \psi_{n-k} \right] ,
\end{equation}
where $g_k(n;\eta)$ is given by
\begin{equation}
  \label{eq:g-func-def}
  g_k(n;\eta) = \kappa (i\eta)^k \sqrt{\frac{(n+k)!}{n!}} \, f_k(n;
  \eta) ,
\end{equation}
and $g_k(n;\eta) \!=\! 0$ for $n \!<\! 0$. For notational simplicity
we will omit here and in the following the time argument of
$\psi_n(t)$ and will only write $\psi_n$.

The time evolution of the populations of the number states, $P_n \!=\!
\psi_n^\ast \psi_n$, are obtained from Eq.~(\ref{eq:diff-eq1}) and its
complex conjugate,
\begin{equation}
  \label{eq:populations1}
  \frac{\partial P_n}{\partial t} = 2 \, {\rm Im} \left[ g_k(n;\eta)
    \, \psi_n^\ast \psi_{n+k} - g_k(n\!-\!k;\eta) \, \psi_{n-k}^\ast
    \psi_n \right] .
\end{equation}
To calculate the second time-derivative of Eq.~(\ref{eq:populations1})
one requires the time-derivatives of combinations of the type
$\psi_n^\ast \psi_{n+k}$ which are given by
\begin{eqnarray}
  \label{eq:correlations}
  \lefteqn{\frac{\partial}{\partial t} \, \psi_n^\ast \psi_{n+k}
    = i g_k^\ast(n;\eta) \left( P_{n+k} - P_n \right)} & & \\
  \nonumber & & \quad + \, i \left[ g_k(n\!-\!k;\eta) \,
    \psi_{n-k}^\ast \psi_{n+k} - g_k(n\!+\!k;\eta) \, \psi_n^\ast
    \psi_{n+2k} \right] .
\end{eqnarray}
Inserting Eq.~(\ref{eq:correlations}) into the time-derivated
Eq.~(\ref{eq:populations1}) the second time-derivative of the number
statistics results as
\begin{eqnarray}
  \label{eq:populations2}
  \frac{\partial^2 P_n}{\partial t^2} & = & 2 \, |g_k(n;\eta)|^2
  \left( P_{n+k} \!-\! P_n \right) \\ \nonumber & - & 2 \,
  |g_k(n\!-\!k;\eta)|^2 \left( P_n \!-\! P_{n-k} \right) \\ \nonumber
  & - & 2 \, {\rm Re} \bigg[ g_k(n;\eta) \, g_k(n\!+\!k;\eta) \,
  \psi_{n+2k}^\ast \psi_n \\ \nonumber & & \quad\quad +
  g_k(n\!-\!k;\eta) \, g_k(n\!-\!2k;\eta) \, \psi_n^\ast \psi_{n-2k}
  \\ \nonumber & & \quad\quad - 2 \, g_k(n;\eta) \, g_k(n\!-\!k;\eta)
  \, \psi_{n+k}^\ast \psi_{n-k} \bigg] .
\end{eqnarray}
We are interested here in the temporal evolution of the mean quantum
number 
\begin{equation}
  \label{eq:mean-n1}
  \frac{d^2 \langle \hat{n}(t) \rangle}{dt^2} = \sum_{n=0}^\infty n \,
  \frac{\partial^2 P_n(t)}{\partial t^2} ,
\end{equation}
which can be calculated with the help of Eq.~(\ref{eq:populations2}).
Here only the first two terms of Eq.~(\ref{eq:populations2})
contribute to the sum in Eq.~(\ref{eq:mean-n1}), whereas the real-part
given in Eq.~(\ref{eq:populations2}) cancels. The resulting equation
of motion for the number statistics reads as
\begin{equation}
  \label{eq:mean-n2}
  \frac{d^2 \! \langle \hat{n}(t) \rangle}{{dt}^2} = 2k
  \sum_{n=0}^\infty \left[ \left| g_k(n;\eta) \right|^2 - \left|
      g_k(n\!-\!k;\eta) \right|^2 \right] P_n(t) .
\end{equation}
Defining the coefficients $F_k(n;\eta)$ by the relation
\begin{equation}
  \label{eq:def-coef}
  \eta^{2k} |\kappa|^2 \, F_k(n;\eta) = k \left[ \left| g_k(n;\eta)
    \right|^2 - \left| g_k(n\!-\!k;\eta) \right|^2 \right] ,
\end{equation}
one obtains for Eq.~(\ref{eq:mean-n2})
\begin{equation}
  \label{eq:mean-n3}
  \frac{d^2 \! \langle \hat{n}(\tau) \rangle}{{d\tau}^2} =
  \sum_{n=0}^\infty F_k(n;\eta) \, P_n(\tau) ,
\end{equation}
with the (dimensionless) scaled time $\tau$ given by
\begin{equation}
  \label{eq:scaled-time}
  \tau = \sqrt{2} \, \eta^k |\kappa| \, t .
\end{equation}
The coefficients $F_k(n;\eta)$ determine the motional dynamics and
especially the existence of a divergence in finite interaction times,
as depicted in Eq.~(\ref{eq:divergence}). From
Eqs.~(\ref{eq:g-func-def}) and (\ref{eq:def-coef}) the coefficients
follow as
\begin{equation}
  \label{eq:F-func-def}
  F_k(n;\eta) = k \left[ \frac{(n\!+\!k)!}{n!} \, f_k^2(n;\eta) -
    \frac{n!}{(n\!-\!k)!}  \, f_k^2(n\!-\!k;\eta) \right] ,
\end{equation}
with $k\!\geq\!1$ and the functions $f_k(n;\eta)$ given by
Eq.~(\ref{eq:k-func-fock}), with $f_k(n;\eta)\!=\!0$ for $n\!<\!0$.

\section{Lamb--Dicke approximation: Exploding solutions}
In the Lamb--Dicke approximation ($\eta\!\to\! 0$) the coefficients
$F_k(n;\eta)$ read as
\begin{equation}
  \label{eq:F-limes}
  F_k(n;0) = \frac{1}{(k\!-\!1)!} \left[ {n+k \choose k} - {n \choose
      k} \right] .
\end{equation}
From Eq.~(\ref{eq:F-limes}) it can be seen that, in general, the
functions $F_k(n;0)$ are polynomials in $n$ of the order $k\!-\!1$,
that is
\begin{equation}
  \label{eq:ld-values}
  F_k(n;0) = \sum_{l=0}^{k-1} a_{kl} \, n^l ,
\end{equation}
with a non-vanishing highest-order coefficient $a_{k,k-1}\!\neq\!0$.
From Eq.~(\ref{eq:F-limes}) it can be seen that the expansion
coefficients are always positive $a_{kl} \!\geq\! 0$. Moreover, the
lowest-order coefficient $a_{k0}$ is non-vanishing, since
\begin{equation}
  \label{eq:lowest-order}
  a_{k0} = F_k(0;0) = \frac{1}{(k-1)!} > 0 .
\end{equation}

The second-order differential equations for the mean excitation number
in the Lamb--Dicke approximation, Eq.~(\ref{eq:mean-n3}) together with
Eq.~(\ref{eq:ld-values}), read as
\begin{equation}
  \label{eq:ld-diffeq}
  \frac{d^2 \langle \hat{n}(\tau) \rangle}{{d\tau}^2} =
  \sum_{l=0}^{k-1} a_{kl} \, \langle \hat{n}^l(\tau) \rangle .
\end{equation}
Note, that the second-derivative~(\ref{eq:ld-diffeq}) is always
positive and non-zero due to the non-vanishing lowest order term
$a_{k0}$, cf. Eq.~(\ref{eq:lowest-order}). For obtaining a lower bound
for the second-derivative we use the relations following from the
Schwarz inequality,
\begin{equation}
  \label{eq:schwarz}
  \langle \hat{n}^l(\tau) \rangle \geq \langle \hat{n}(\tau) \rangle^l
  , \quad l=0,1,2,\ldots
\end{equation}
Due to the positiveness of the coefficients $a_{kl}$ a lower bound of
the right-hand side of Eq.~(\ref{eq:ld-diffeq}) follows by using
Eq.~(\ref{eq:schwarz}),
\begin{equation}
  \label{eq:bound1}
  \frac{d^2 \langle \hat{n}(\tau) \rangle}{{d\tau}^2} =
  \sum_{l=0}^{k-1} a_{kl} \langle \hat{n}^l(\tau) \rangle \geq
  \sum_{l=0}^{k-1} a_{kl} \langle \hat{n}(\tau) \rangle^l > 0 .
\end{equation}
From the formal solution of Eq.~(\ref{eq:ld-diffeq}) and the
application of Eq.~(\ref{eq:bound1}) one obtains the inequality
\begin{eqnarray}
  \label{eq:formal-solution}
  \langle \hat{n}(\tau) \rangle & = & \bar{n}_0 + \bar{n}_0' \, \tau +
  \int_0^\tau \!\!  d\tau' \int_0^{\tau'}
  \!\! d\tau'' \, \frac{d^2 \langle \hat{n}(\tau'') \rangle}{{d\tau}^2} \\
  \nonumber & \geq & \bar{n}_0 + \bar{n}_0' \, \tau + \int_0^\tau \!\!
  d\tau' \int_0^{\tau'} \!\!  d\tau'' \sum_{l=0}^{k-1} a_{kl} \,
  \langle \hat{n}(\tau'') \rangle^l ,
\end{eqnarray}
with the initial conditions 
\begin{equation}
  \label{eq:initial-conditions}
  \bar{n}_0 = \left. \langle \hat{n}(\tau) \rangle \right|_{\tau=0} ,
  \quad \bar{n}_0' = \left. \frac{d\langle \hat{n}(\tau)
      \rangle}{d\tau} \right|_{\tau=0} ,
\end{equation}
where we have chosen, without loss of generality, the initial time to
be $\tau\!=\!0$. From Eq.~(\ref{eq:formal-solution}) it follows, that
the solution $N_{\rm lb}(\tau)$ of the second-order differential
equation
\begin{equation}
  \label{eq:bound2}
  \frac{d^2 N_{\rm lb}(\tau)}{{d\tau}^2} = \sum_{l=0}^{k-1} a_{kl} \,
  N_{\rm lb}^l(\tau) .
\end{equation}
obeys the relation
\begin{equation}
  \label{eq:low-bound}
  N_{\rm lb}(\tau) \leq \langle \hat{n}(\tau) \rangle , \quad (\tau
  \geq 0) ,
\end{equation}
for $\tau\!\geq\!0$ and identical initial conditions $N_{\rm lb}(0)
\!=\!  \bar{n}_0$ and $N_{\rm lb}'(0) \!=\!  \bar{n}_0'$.  That is,
$N_{\rm lb}(\tau)$ represents a lower-bound (lb) for the solution
$\langle \hat{n}(\tau) \rangle$. We will show in the following, that
for $k \!\geq\! 3$ the lower-bound solution $N_{\rm lb}(\tau)$ may
diverge for finite interaction times, so that it is proved, that the
correct solution $\langle \hat{n}(\tau) \rangle$ also diverges.

We obtain a first-order differential equation by considering the first
derivative $N_{\rm lb}'(\tau) \!=\! d N_{\rm lb}(\tau) / d\tau$,
\begin{equation}
  \label{eq:first-order}
  \frac{dN_{\rm lb}'}{d\tau} = \frac{dN_{\rm lb}'}{dN_{\rm lb}} \,
  \frac{dN_{\rm lb}}{d\tau} = \frac{dN_{\rm lb}'}{dN_{\rm lb}} \,
  N_{\rm lb}' = \sum_{l=0}^{k-1} a_{kl} \, N_{\rm lb}^l .
\end{equation}
The last equality in Eq.~(\ref{eq:first-order}) can then be easily
solved by integration of
\begin{equation}
  \label{eq:integr-examp}
  N_{\rm lb}' \, dN_{\rm lb}' = \sum_{l=0}^{k-1} a_{kl} \, N_{\rm
    lb}^l \, dN_{\rm lb} ,
\end{equation}
 and one obtains
\begin{equation}
  \label{eq:integrated}
  \left[ \frac{dN_{\rm lb}(\tau)}{d\tau} \right]^2 = \bar{n}^{\prime
    2}_0 + \sum_{l=1}^k b_{kl} \left[ N_{\rm lb}^l(\tau) - \bar{n}^l_0
  \right] ,
\end{equation}
with the coefficients $b_{kl} \!=\! 2 \, a_{k,l-1} / l \!\geq\! 0$ and
$b_{k1} \!\neq\! 0$, $b_{kk} \!\neq\! 0$.

To demonstrate the unphysical properties of the Hamiltonian
$\hat{H}_k^{\rm (LD)}$ for $k\!\geq\!3$ in the Lamb--Dicke
approximation~(\ref{eq:k-hamilton-limes}), it is sufficient to prove
the unphysical behaviour for one physically reasonable initial
condition. For the special cases $k\!=\!3,4$ this has been already
explicitly shown in Refs.~\cite{elyutin,hillery}, here we want to show
the unphysical behaviour, in a general way, for all $k\!\geq\!3$.  For
the atom initially (at $\tau\!=\!0$) in its vibrational ground-state,
$|\psi(0)\rangle \!=\! |0\rangle$, the initial conditions are
$\bar{n}_0 \!=\!  \bar{n}_0' \!=\! 0$ [the latter can be seen from
Eq.~(\ref{eq:populations1}) with $\psi_n(0)\!=\!\delta_{n,0}$] and the
differential equation~(\ref{eq:integrated}) reduces to
\begin{equation}
  \label{eq:integrated1b}
  \frac{dN_{\rm lb}(\tau)}{d\tau} = \left[ \sum_{l=1}^k b_{kl} \,
    N_{\rm lb}^l(\tau) \right]^{1/2} .
\end{equation}
Here we have chosen the positive square-root since for vanishing
initial {\em velocity}, $\bar{n}_0'\!=\!0$, and always positive {\em
  acceleration}, $d^2N_{\rm lb}(\tau) / d\tau^2 \!>\! 0$ [cf.
Eqs.~(\ref{eq:bound1}) and (\ref{eq:bound2})], the {\em velocity} at
time $\tau \!>\!0$ has to be positive, $dN_{\rm lb}(\tau)/d\tau \!>\!
0$.  Eq.~(\ref{eq:integrated1b}) can then be integrated from the
finite time $\tau_1\!>\!0$ to $\tau_2\!\geq\!\tau_1$ which gives the
relation
\begin{equation}
  \label{eq:integrated2}
  \tau_2 - \tau_1 = \int_{N_{\rm lb}(\tau_1)}^{N_{\rm lb}(\tau_2)}
  \!\!  \frac{dn}{\sqrt{ b_{kk} \, n^k + \ldots + b_{k2} \, n^2 +
      b_{k1} \, n^1}} .
\end{equation}
Since the {\em velocity} $dN_{\rm lb}(\tau)/d\tau$ is always positive
and non-vanishing for $\tau \!>\!0$, it is clear, that $N_{\rm
  lb}(\tau_1) \!>\!  \bar{n}_0 \!=\! 0$. Therefore the integration
starts with a positive and non-vanishing value of the excitation,
$N_{\rm lb}(\tau_1) \!>\! 0$, that has been attained after the
interaction time $\tau_1$.

Now we are interested in the further evolution in the time interval
$\tau_2\!-\!\tau_1$. In particular, we are looking for that time
interval $\Delta\tau_\infty \!=\! \tau_2\!-\!\tau_1$ for which the
excitation number $N_{\rm lb}(\tau_2)$ attains an infinite value
$N_{\rm lb}(\tau_2)\!\to\!\infty$. By taking only the highest-order
term in the square-root of Eq.~(\ref{eq:integrated2}) we obtain an
upper bound for $\Delta\tau_\infty$
\begin{equation}
  \label{eq:integrated3}
  \Delta\tau_\infty \leq \int_{N_{\rm lb}(\tau_1)}^\infty \!\!
  \frac{dn}{\sqrt{ b_{kk} \, n^k }} = \left\{ \begin{array}{cl} \infty
      , & (k\!=\!1,2) , \\[1ex] \frac{2}{k-2} \frac{1}{\sqrt{b_{kk} \,
          N_{\rm lb}^{k-2}(\tau_1)}} , & (k \!\geq\! 3) . \end{array}
  \right.
\end{equation}
This result reveals that we obtain a finite value of
$\Delta\tau_\infty$ for $k\!\geq\!3$. That is, after attaining the
finite (non-vanishing) excitation $N_{\rm lb}(\tau_1)$ after the
interaction time $\tau_1$, the solution $N_{\rm lb}(\tau)$ of
Eq.~(\ref{eq:bound2}) diverges already after the finite time interval
$\Delta\tau_\infty$, according to Eq.~(\ref{eq:integrated3}).
Concluding, for $k\!\geq\!3$ the solution of Eq.~(\ref{eq:ld-diffeq})
[$\langle \hat{n}(\tau_2) \rangle \!\geq\! N_{\rm lb}(\tau_2)$] will
also diverge at a certain finite interaction time $\tau_2 \!\leq\!
\tau_1 \!+\! \Delta\tau_\infty$. For $k\!=\!1,2$ no upper bound for the
interaction time can be given and it can be seen by direct integration
of Eq.~(\ref{eq:ld-diffeq}) that the mean excitation does not diverge
in a finite interaction time.

\section{Overlap of matter and light waves: Regular behaviour for large
  excitations} In the preceding section it has been shown that in the
Lamb--Dicke approximation the mean motional excitation number diverges
in a finite interaction time for the cases $k\!\geq\!3$. In this
section we will proof, that the exact
Hamiltonian~(\ref{eq:k-hamiltonian}), e.g.  without the Lamb--Dicke
approximation, does not exhibit such a divergence problem. This is due
to the overlap of matter and light waves described by the nonlinear
operator functions~(\ref{eq:k-function}).  They lead to an
excitation-dependent coupling strength which suppresses the unbounded
increase of the mean excitation.

\subsection{Proof of the regular behaviour}
To prove the regular behaviour of the dynamics of the system described
by Eqs.~(\ref{eq:mean-n3}) and (\ref{eq:F-func-def}) we may consider
the following situation: \\
(a) If the mean quantum number would diverge, we would be operating in
a regime of very large quantum numbers $n$. Therefore we are allowed
to use an asymptotic expansion of the coefficients
$F_k(n;\eta)$ for large $n$. \\
(b) Since for $k\!=\!1,2$ we know that in the Lamb--Dicke
approximation [described by $F_{1,2}(n;0)$] the dynamics does not
exhibit a divergence in finite interaction times, it is sufficient to
show that the asymptotic expansion of $F_k(n;\eta)$ has an upper bound
leading to a dynamics which is at least as convergent as for
$F_{1,2}(n;0)$,
\begin{equation}
  \label{eq:to-show}
  F_k(n;\eta) \leq F_{1,2}(n;0) , \quad (n\gg 1) .
\end{equation}
Then the {\em acceleration} $d^2\langle\hat{n}(\tau)\rangle/d\tau^2$
is always smaller than those for the well-behaved cases and a
divergence in finite times cannot exist, regardless of the initial
motional quantum state chosen.

We start by expressing the function $F_k(n;\eta)$ given in
Eq.~(\ref{eq:F-func-def}) in terms of Laguerre polynomials by using
Eq.~(\ref{eq:k-func-fock}),
\begin{eqnarray}
  \label{eq:F-func-laguerre}
  F_k(n;\eta) & = & \bigg\{ \frac{n!}{(n+k)!} \left[ L_n^{(k)}(\eta^2)
  \right]^2 \\ \nonumber & & \quad - \frac{(n-k)!}{n!} \left[
    L_{n-k}^{(k)}(\eta^2) \right]^2 \bigg\} \, e^{-\eta^2/2} .
\end{eqnarray}
While the first (positive) term in Eq.~(\ref{eq:F-func-laguerre})
describes the transition to higher-lying states
$|n\rangle\!\to\!|n\!+\!k\rangle$, the second (negative) term
describes transitions to lower-lying states
$|n\rangle\!\to\!|n\!-\!k\rangle$, leading to a decrease of the {\em
  acceleration}. An upper bound for $F_k(n;\eta)$, which determines
the maximum {\em acceleration}, is therefore given by neglecting the
transitions to lower-lying states (which do not cause a divergent
behaviour)
\begin{equation}
  \label{eq:up-bound}
  F_k(n;\eta) \leq \frac{n!}{(n+k)!} \left[ L_n^{(k)}(\eta^2)
  \right]^2 \, e^{-\eta^2/2} .
\end{equation}
Using the relation between the Laguerre polynomials and the confluent
hypergeometric (Kummer's) function $M(a,b;x)$~\cite{abramowitz-stegun}
\begin{equation}
  \label{eq:hyper-laguerre}
  L_n^{(k)}(x) = {n+k \choose n} \, M(-n,k+1;x) ,
\end{equation}
one arrives at the inequality for $F_k(n;\eta)$
\begin{equation}
  \label{eq:F-func-hyper}
  F_k(n;\eta) \leq \frac{1}{k!}  {n+k \choose k} \, M^2(-n,k+1;\eta^2)
  \, e^{-\eta^2/2} .
\end{equation}
An asymptotic expansion of the confluent hypergeometric function
$M(a,b;x)$ for $a\!\to\!-\infty$, bounded $b$, and real-valued $x$ is
given by~\cite{abramowitz-stegun},
\begin{eqnarray}
  \label{eq:M-asymptotic}
  M(a,b;x) & \sim & \frac{\Gamma(b)}{\sqrt{\pi}} \, e^{\frac{1}{2} x}
  \left[ \left(\frac{b}{2} - a \right) x
  \right]^{\frac{1}{4}-\frac{1}{2} b} \\ \nonumber & & \times \,
  \cos\left[ \sqrt{ (2b-4a) x} - \frac{1}{2} b \pi + \frac{1}{4} \pi
  \right] .
\end{eqnarray}
Thus, for large numbers $n$ the inequality reads in its asymptotic
form
\begin{eqnarray}
  \label{eq:F-asymptotic}
  F_k(n;\eta) & \leq & \frac{1}{\pi} \frac{(n\!+\!k)!}{n!} \left[
    \eta^2 \left( n \!+\!
      \frac{1\!+\!k}{2} \right) \right]^{-k-\frac{1}{2}} \nonumber \\
  & \times & \cos^2\left[ 2 \eta \sqrt{ n \!+\!  \frac{1\!+\!k}{2}} -
    \frac{1}{2} k \pi - \frac{1}{4}\pi \right] \, e^{\eta^2/2} .
\end{eqnarray}
Therefore, the maximum value of the right-hand side of
Eq.~(\ref{eq:F-asymptotic}) can be estimated by taking the squared
cosines to be unity,
\begin{equation}
  \label{eq:F-bound}
  F_k(n;\eta) \leq \frac{(n+k)!}{\pi \, n!}  \left[ \eta^2 \left( n
      \!+\!  \frac{1\!+\!k}{2} \right) \right]^{-k-\frac{1}{2}} \,
  e^{\eta^2/2} .
\end{equation}
Moreover, the expression~(\ref{eq:F-bound}) can be further estimated
by the following relation
\begin{eqnarray}
  \label{eq:fac-relation}
  \frac{(n+k)!}{n!} & = & (n+k) (n+k-1) \ldots (n+1) \nonumber \\ &
  \leq & (n+k)^k ,
\end{eqnarray}
which gives one a further simplification,
\begin{equation}
  \label{eq:F-bound2}
  F_k(n;\eta) \leq \frac{1}{\pi} \, e^{\eta^2/2} \frac{1}{\sqrt{
      \eta^2 \left( n \!+\! \frac{1\!+\!k}{2} \right)}} \left[
    \frac{(n+k)}{\eta^2 \left( n \!+\! \frac{1\!+\!k}{2} \right)}
  \right]^{k} .
\end{equation}
For the range of large numbers $n$, we are considering here, the
function therefore has the following upper bound,
\begin{equation}
  \label{eq:F-bound3}
  F_k(n;\eta) \leq \frac{1}{\pi} \, \frac{e^{\eta^2/2}}{\eta^{2k+1}}
  \, \frac{1}{\sqrt{ n }} ,
\end{equation}
that is, for large numbers $n$ the upper bound of the function
$F_k(n;\eta)$ decays as $1/\sqrt{n}$. It therefore can be further
estimated by a simple constant $C_k(\eta)$,
\begin{equation}
  \label{eq:F-bound4}
  F_k(n;\eta) \leq C_k(\eta) , \quad C(\eta) = \frac{1}{\pi} \,
  \frac{e^{\eta^2/2}}{\eta^{2k+1}} .
\end{equation}
The resulting differential equation for the upper bound (ub) $N_{\rm
  ub}(\tau)$ of mean quantum number reduces then for possibly large
numbers $n$ to
\begin{equation}
  \label{eq:diff-eq-bound}
  \frac{d^2N_{\rm ub}(\tau)}{d\tau^2} = C_k(\eta) .
\end{equation}
Reconsidering the formal solution~(\ref{eq:formal-solution}) and the
upper-bound {\em acceleration}~(\ref{eq:F-bound4}) it becomes clear
that $N_{\rm ub}(\tau)$ indeed is an upper bound for the exact mean
excitation number,
\begin{equation}
  N_{\rm ub}(\tau) \geq \langle \hat{n}(\tau) \rangle ,
\end{equation}
for identically chosen initial conditions $N_{\rm
  ub}(0)\!=\!\bar{n}_0$, $N_{\rm ub}'(0)\!=\!\bar{n}_0'$ and large
excitations, $\langle \hat{n}(\tau) \rangle \!\gg\! 1$.
Eq.~(\ref{eq:diff-eq-bound}) states, that the mean excitation number
does not diverge in finite time, since the differential equation for
large values of $n$ leads to a behaviour which is as convergent as in
the case of $k\!=\!1$ in the Lamb--Dicke limit where $F_1(n;0) \!=\! 1
\!=\!  \mbox{const}$. That is, the upper-bound solution of
Eq.~(\ref{eq:diff-eq-bound}), $N_{\rm ub}(\tau)$, which can be
obtained by direct integration,
\begin{equation}
  \label{eq:k1-solution}
  N_{\rm ub}(\tau) = \bar{n}_0 + \bar{n}_0' \, \tau + \frac{1}{2} \,
  C_k(\eta) \, \tau^2 ,
\end{equation}
does not diverge for finite interaction times $\tau$. In conclusion,
it has been proved that the mean motional excitation number resulting
from the full Hamiltonian~(\ref{eq:k-hamiltonian}) does not diverge
for finite interaction times.

\subsection{Numerical examples}
\begin{figure}
  \begin{center}
    \psfig{file=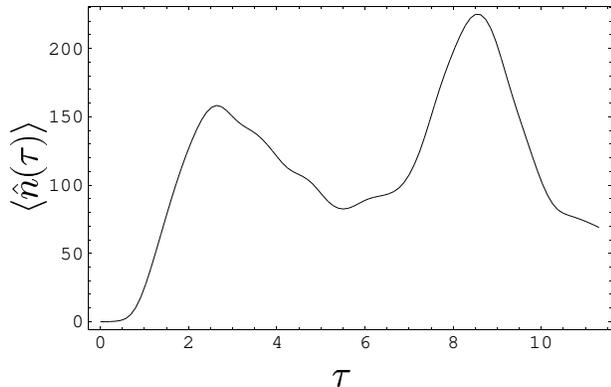,scale=0.65}
    \caption{Exact time evolution of the mean motional excitation
      number $\langle \hat{n}(\tau) \rangle$ for $k\!=\!3$ and
      Lamb--Dicke parameter $\eta\!=\!0.2$, as a function of the
      scaled time $\tau$ given in Eq.~(\ref{eq:scaled-time}).}
    \label{fig:meann}
  \end{center}
\end{figure}

As an example, we show in Fig.~\ref{fig:meann} the exact time
evolution of the mean motional excitation number $\langle
\hat{n}(\tau) \rangle$ for $k\!=\!3$ and Lamb--Dicke parameter $\eta
\!=\! 0.2$. It clearly shows, that instead of diverging in a finite
interaction time as would be expected in the Lamb--Dicke
approximation, the mean excitation number exhibits an oscillatory
behaviour. This is due to the destructive overlap of matter and light
waves leading to a decoupling of the atomic motion from the laser
fields for certain excitation amplitudes.

\begin{figure*}
  \begin{center}
    \vspace*{-1.15cm}
    \hspace*{-2cm}
    \input{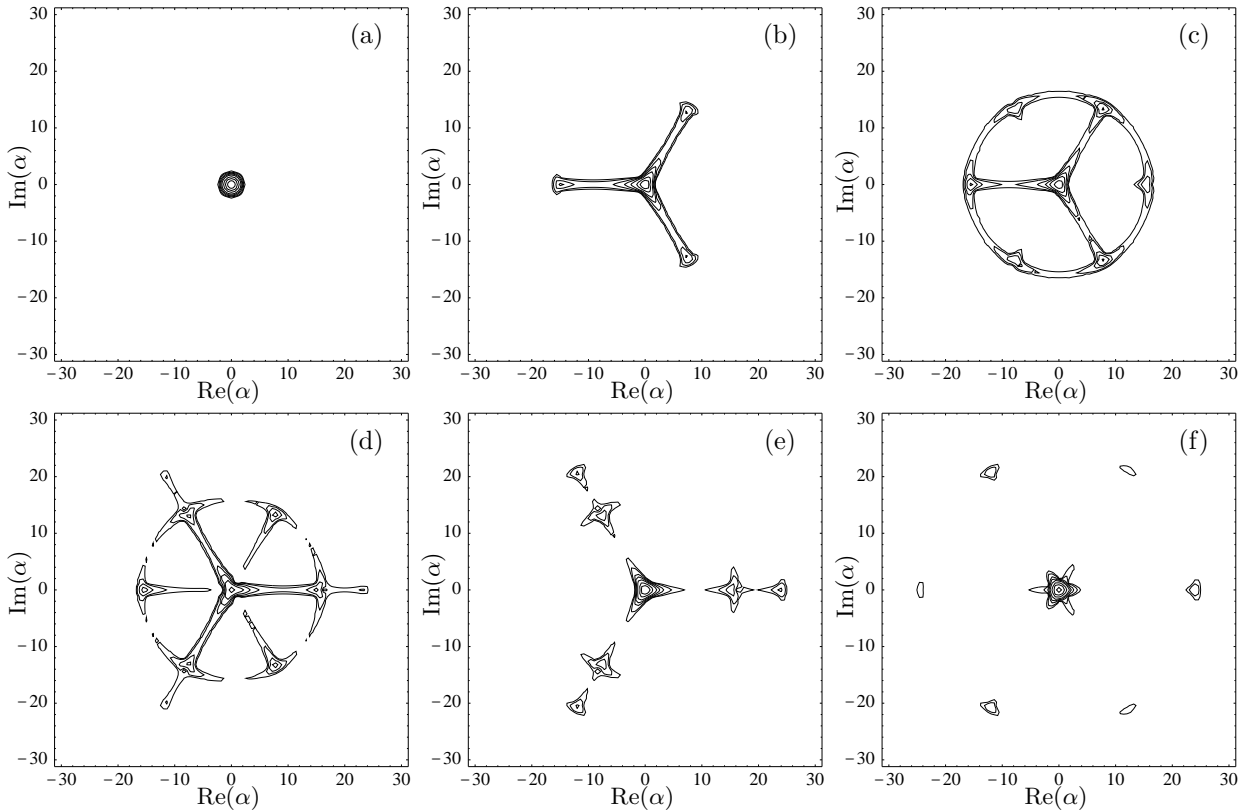}
    \vspace*{2ex}
    \caption{Time evolution of the Husimi Q function for
      an initial motional ground state under the influence of the
      three-quantum coupling ($k\!=\!3$). The Lamb--Dicke parameter
      has been chosen $\eta\!=\!0.2$ and the scaled times $\tau$ are:
      0 (a), 1.14 (b), 2.29 (c), 3.44 (d), 4.59 (e), and 5.74 (f).
      Note the formation of a ``star'' followed by a ring from which
      further structure grows.}
    \label{fig:qfunction}
  \end{center}
\end{figure*}

To gain more insight into the distribution of the quantum state in
phase space, we show in Fig.~\ref{fig:qfunction} the time evolution of
the {\em Q} function for the three-quantum coupling ($k\!=\!3$) and
for $\eta\!=\!0.2$.  It can be seen that the dynamics is strongly
modified by the occurrence of the circles of vanishing coupling
strengths. In contrast to the dynamics in the Lamb--Dicke
approximation, where the ``star''-like structure would be extended to
infinitely large phase-space amplitudes, the extension of the ``star''
structure is halted at the first circle of vanishing coupling. Parts
of the phase-space distribution are smoothed over the circle. For
those components of the distribution that accumulate a phase shift of
$\pi/3$ relative to the initial ``star'' structure, the
Hamiltonian~(\ref{eq:k-hamiltonian}) effectively exhibits a change of
sign accompanied by a reversal of the time evolution. Consequently,
those components of the quantum state are moving back towards the
origin of phase space~\cite{final-remark}.  This effect explains the
decrease of the mean motional excitation number as seen in
Fig.~\ref{fig:meann}. Note that the distribution in
Fig.~\ref{fig:qfunction}(f) for time $\tau\!=\!5.74$ corresponds to a
local minimum of $\langle \hat{n}(\tau) \rangle$ in
Fig.~\ref{fig:meann}.  Obviously, there are some components of the
phase-space distribution which cross the barrier.  However, because of
the existence of further barriers at approximately equidistant radii,
the explosive dynamics occurring in the Lamb--Dicke approximation and
also in the optical parametric approximation does not occur.

\section{Summary and conclusions}
In summary it has been shown that for a trapped atom which is driven
by Raman-laser fields, in the Lamb--Dicke approximation a behaviour
appears which is analogous to the case of $k$-photon down conversion
in nonlinear optics. A divergent behaviour of the mean motional
excitation number after finite interaction times occurs for
higher-order quantum couplings with $k\!\geq\!3$, similar to the
situation for the parametric approximation in nonlinear optics. We
have discussed these divergences within a single unified framework for
all orders $k\!\geq\!3$. Moreover, it has been argued that the
Lamb--Dicke approximation, which is only valid for well-localised
atoms, is not consistent with the occurrence of large (or even
diverging) mean excitations.

To overcome the divergent behaviour, one has to treat the full problem
without the Lamb--Dicke approximation.  This includes the correct
description of the laser-induced momentum transfer onto the
center-of-mass of the trapped atom. These are described by a nonlinear
operator function, which plays an essential role for the dynamics of
the motional quantum state of the atom. By using an asymptotic
expansion, we have proved that the correct description of the recoil
effects widely modifies the dynamics for large excitations and
prevents the mean excitation number from exploding for finite
interaction times.  That is, the full problem leads to a regular
dynamics where the energy of the motional degree of freedom does not
unphysically diverge. On the other hand, the Lamb--Dicke approximation
fails for these types of couplings, as does the parametric
approximation in nonlinear optics.  Whereas in nonlinear optics the
divergence problem arises from the neglection of the pump-mode
depletion and entanglement of the involved field modes, in the case of
a trapped atom the unappropriate treatment of the recoil effects in
the Lamb--Dicke approximation leads to the unphysical behaviour.

\section*{Acknowledgements}
This work was supported by the Deutsche Forschungsgemeinschaft, by the
UK Engineering and Physical Sciences Research Council, and by the
European Union.

\begin{references}
\bibitem{k-remark} For $k\!=\!0$ one obtains a zero-quantum coupling,
  where no vibrational quanta are created or annihilated in the motion
  of the trapped atom. For this case only the excitation-dependent
  operator function $f_0(\hat{n};\eta)$ acts on the motional degree of
  freedom and leads to phase-shift and dispersive effects, for more
  details see Ref.~\cite{counterpart}. We exclude this case here,
  since we are mainly interested in the particular situation for
  $k\!\geq\!3$.
\end{references}
\end{document}